\documentclass[%
 aip,
 amsmath,amssymb,
 reprint,%
]{revtex4-1}

\usepackage{graphicx}
\usepackage{dcolumn}
\usepackage{bm}

\usepackage[utf8]{inputenc}
\usepackage[T1]{fontenc}
\usepackage{mathptmx}
\usepackage{etoolbox}

\makeatletter
\def\@email#1#2{%
 \endgroup
 \patchcmd{\titleblock@produce}
  {\frontmatter@RRAPformat}
  {\frontmatter@RRAPformat{\produce@RRAP{*#1\href{mailto:#2}{#2}}}\frontmatter@RRAPformat}
  {}{}
}%
\makeatother
\begin{document}

\preprint{AIP/123-QED}

\title[Cell aspect ratio important for bacteria]{Cell aspect ratio important for bacterial rheotaxis}
\author{Jiachen Zhong}
\affiliation{Complex Systems Division, Beijing Computational Science Research Center, Beijing 100193, China}
\author{Xinliang Xu}
\email{xinliang@hainanu.edu.cn}
\affiliation{School of Physics and Optoelectronic Engineering, Hainan University, Haikou 570228, China}
\affiliation{Complex Systems Division, Beijing Computational Science Research Center, Beijing 100193, China}
\affiliation{Department of Physics, Beijing Normal University, Beijing 100875, China}

\date{\today}

\begin{abstract}
When placed in flows with local shear, flagellated bacteria commonly display reorientations towards the local vorticity direction, a chirality-induced rheotactic behavior of great importance for many biological functions. As the observed reorientational dynamics arises from the interplay between the Jeffery dynamics controlled by the cell aspect ratio and the chirality-induced reorientation controlled by the chiral strength, in this work we quantitatively study how cell aspect ratio can become a key factor in tuning the reorientational dynamics, through numerical and theoretical analysis. Our results reveal that, for sphere-like bacteria the rheotactic behavior sensitively depends on the cell aspect ratio and totally disappears in the limiting case with cell aspect ratio being 1, in very good agreement with our predicted analytic functional form. For needle-like bacteria with cell aspect ratio much larger than 1, the rheotactic behavior becomes insensitive to cell aspect ratio, in agreement with previous experimental observations.
\end{abstract}

\maketitle

\section{\label{sec:level1}Introduction}

For microswimmers, being able to respond to flows in environment is vital. Common examples of such rheotactic behaviors include upstream navigation of sperms \cite{Bretherton1961, miki2013} or bacteria \cite{hill2007} or man-made microswimmers \cite{Brosseau2019}, shear-induced trapping \cite{durham2009, rusconi2014, barry2015}, and swinging for {\it E. coli} bacteria in Poiseuille flow \cite{zottl2012, junot2019}. Among these behaviors, the reorientation of cells in response to external flow is intensely studied due to its importance for environmental and medical issues such as drug delivery \cite{simone2008}, bio-contamination \cite{costerton1999, von2005, bain2014}, and sorting/analysis of cells \cite{brown2000, hatzenpichler2016}. It is widely believed that this reorientational behavior can be achieved by the intrinsic chirality of microorganisms, e.g., the helical structure of spirochetes \cite{nakamura2020} or bacterial flagella \cite{lauga2016}. When coupled with local shear flows, this intrinsic chirality breaks symmetry and leads to cell translation in the positive/negative vorticity direction based on the handedness \cite{zottl2023, marcos2009, meinhardt2012, ishimoto2020}. For {\it E. coli} bacteria with a non-chiral component of bacterial body and a chiral component of flagella bundle, such coupling between local shear and the chiral component leads to a rheotactic torque that reorients the cell towards the positive/negative vorticity direction \cite{marcos2012, jing2020}. This chirality-induced rheotaxis, i.e. the reorientation in response to local flow gradient, for flagellated bacterial like {\it E. coli} can be formally characterized by a phenomenological parameter, the chiral strength factor $\nu_c$, which is an intrinsic property only related to the specific shape of the chiral flagella bundle and the size of bacterial body \cite{mathijssen2019}. Combining this simple characterization of chirality-induced reorientation with the standard Jeffery dynamics \cite{jeffery1922}, a previous study showed that the experimental observed bacterial orientational dynamics in a poiseuille flow can be well reproduced numerically \cite{jing2020}. However, if/how such chirality-induced reorientations can be affected by the cell aspect ratio $\alpha$, which is an independent parameter controlling the Jeffery dynamics, remains not well understood. In this study we follow the same numerical procedure used in previous work \cite{jing2020} and quantitatively investigate the interplay between Jeffery dynamics and chirality-induced reorientations, controlled by independent parameters $\alpha$ and $\nu_c$, respectively. Our results show that the rheotactic dynamics depends strongly on cell aspect ratio $\alpha$ when $\alpha\approx 1$, and becomes insensitive to $\alpha$ when $\alpha \gg 1$. More importantly, the reorientational dynamics becomes qualitatively different in the limiting case with $\alpha=1$: cells no longer reorient towards the flow vorticity direction and thus show no rheotactic behavior for all $\nu_c$.

\section{Model and simulation methods}
\begin{figure}
\includegraphics[width=1\linewidth]{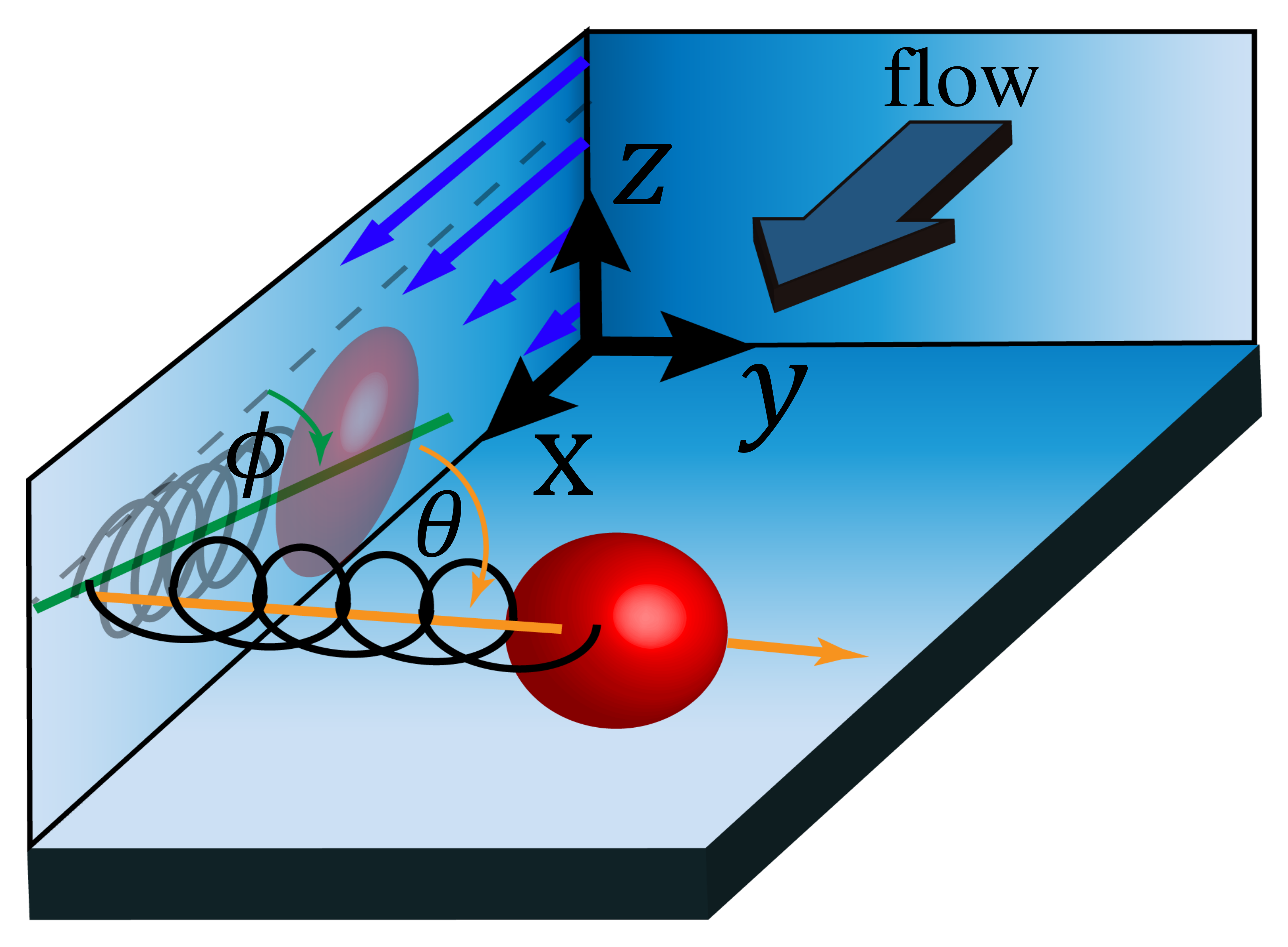}
\caption{Model system. We have one bacterium (red sphere for bacterial body and black helix for flagellar bundle) swimming in water with an external flow in the $x$ direction, which has a gradient in the $z$ direction. The bacterial orientation is defined through the azimuthal angle $\phi$ and polar angle $\theta$.}
\label{Model}
\end{figure}

For a quantitative study of the chirality-induced rheotaxis of the bacterium, we simulate the bacterial motion following a previous classical study \cite{jing2020}. Neglecting the interactions between bacteria, the model system in this study consists only one {\it E. coli} bacterium in an infinitely large space filled by water, and the results are in good agreement with experiments of dilute bacterial systems \cite{jing2020}. When subject to external flow, the bacterial motion is just
\begin{eqnarray}
\dfrac{d \bm{r}}{dt}&=&u \bm{p}+ \bm{v} + \sqrt{2D_t}\xi \label{eomTra} \\
\dfrac{d \bm{p}}{dt}&=&(\bm{\Omega}^J+\bm{\Omega}^C+\sqrt{2 D_r}\xi) \times \bm{p} \label{eomOri}
\end{eqnarray}
where $D_t$ and $D_r$ are the translational and orientational diffusion coefficients, respectively, $\xi$ is the Gaussian white noise with zero mean and unit standard deviation, $\bm{p}$ is an unit vector characterizing bacterial orientation, $u=25 \mathrm{\mu m/s}$ is bacterial swimming speed arising from self propulsion, $\bm{v}$ is the external flow velocity at bacterial location $r$, $\bm{\Omega}^J$ is the Jeffery reorientation rate, and $\bm{\Omega}^C$ is the chirality-induced reorientation rate. Here we study a simple case where the external flow is only in one direction, namely $\bm{v}=v_x \hat{\bm{x}}$ (Fig.\,\ref{Model}), which is very common in experiments \cite{jing2020}. Accordingly, bacterial orientation can be described in terms of polar angle $\theta$ between bacterium and the $xz$ plane and the azimuthal angle $\phi$, as $\bm{p}\equiv (-\cos{\theta} \cos{\phi} \hat{\bm{x}}, \sin{\theta} \hat{\bm{y}}, -\cos{\theta}\sin{\phi}\hat{\bm{z}})$ as illustrated in Fig.\,\ref{Model}. For bacterial orientational dynamics, it is shown that the external flow can be fully accounted for using a single parameter $\dot{\gamma}\equiv \partial v_x/\partial z$, and the deterministic part of the orientational dynamics in Eq.~(\ref{eomOri}) can be rewritten as \cite{mathijssen2019,raible2004,dhont1996,jeffery1922,jing2020,zottl2013}:
\begin{eqnarray}
\dot{\phi}&=&\dfrac{\dot{\gamma}}{2}\left(G\cos {2\phi}-1\right)+\nu_c\dot{\gamma}\sin {2\phi} \sin \theta \label{azi} \\
\dot{\theta}&=&-G\dfrac{\dot{\gamma}}{4}\sin {2\phi} \sin {2\theta}+\nu_c\dot{\gamma}\cos {2\phi} \cos \theta \label{polar}
\end{eqnarray}
where we have used the Bretherton shape factor $G$, which is a simple function of bacterial aspect ratio $\alpha \geq 1$ in the form of $G\equiv (\alpha^2-1)/(\alpha^2+1)\in [0,1]$; and the chiral strength factor $\nu_c$, which is an independent parameter arising from the specific dimension of both the bacterial body and the the left-handed chiral flagella bundle \cite{marcos2012,mathijssen2019}. Detailed relation between $\nu_c$ and bacterial dimensions is available in appendix \ref{appendix A}.

In this study we treat $G$ (or equivalently the aspect ratio $\alpha$) and $\nu_c$ as two independent tuning parameters, and study bacterial rheotaxis in response to simple shearing flows where $\dot{\gamma}$ is a constant. In section III we perform simulations for a specific $\dot{\gamma}$. The data helped lead us to an analytic description of bacterial rheotaxis as a simple function of $\alpha$ and $\nu_c$, which is then verified by more simulations for simple shearing flows with other different choices of $\dot{\gamma}$. In accordance to existing studies \cite{jing2020}, here for bacteria with fixed swimming speed $u$ the bacterial rheotaxis is characterized by the mean bacterial orientation in the vorticity direction of the external flow, i.e., $\langle p_y \rangle$. Since $\dot{\gamma}$ is constant for a simple shearing flow, the orientational dynamics is independent of bacterial location and thus $\langle p_y \rangle$ can be obtained without solving Eq.~(\ref{eomTra}). In this case for each choice of $\{\alpha,\nu_c\}$ we perform 4,000 independent runs with random initial orientation, i.e. random $\theta\in[0,\pi]$ and random $\phi\in[-\pi,\pi]$, where in each run we only simulate the orientational dynamics based on Eq.~(\ref{eomOri}). In all our simulations, we use $1\mu m$ and $1s$ as the units for length and time, respectively. And the time step $dt$ is chosen to keep $dt\times Pe=0.2$ a constant, where rotational Péclet numbers $Pe \equiv \dot{\gamma}/D_r$ all satisfy $Pe \gg 1$ in our studies.

\begin{figure}
\includegraphics[width=1\linewidth]{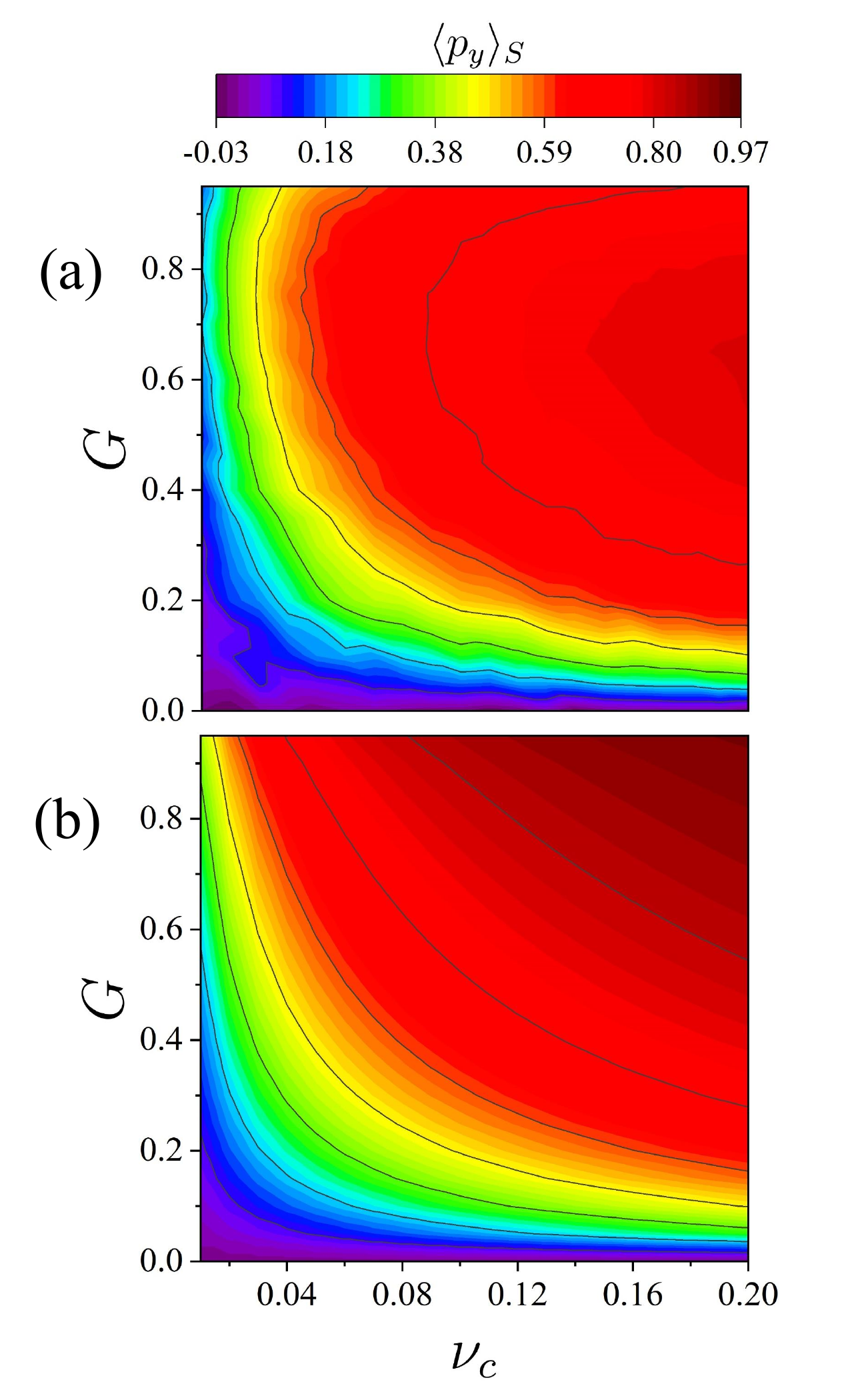}
\caption{Contour plots of $\langle p_y \rangle_S$ from simulations (a) and theoretical predicted by Eq.~(\ref{SteadyState}) (b), in terms of Bretherton shape factor $G$ and chiral strength factor $\nu_c$, for a simple shearing flow with $\dot{\gamma}=10$.}
\label{Steady}
\end{figure}

\section{numerical results}
For simple shearing flow with $\dot{\gamma}=10s^{-1}$, we simulated bacterial orientational dynamics based on Eq.~(\ref{eomOri}) at different choices of $\{\alpha, \nu_c\}$ within a range that matches well with experiments \cite{mathijssen2019,jing2020}. For each choice of $\{\alpha, \nu_c\}$, 4,000 independent runs are performed and the average provides us system chirality-induced rheotaxis characterized by $\langle p_y \rangle$, which reaches to a steady state $\langle p_y \rangle_S$ at long time limit. Fig.\,\ref{Steady}a shows simulation results of $\langle p_y \rangle_S$ as a function of $G\equiv(\alpha^2-1)/(\alpha^2+1)$ and $\nu_c$ in a contour map, where bacterial rheotaxis at all levels from being weak ($\langle p_y \rangle_S\approx 0$) to being strong ($\langle p_y \rangle_S\approx 1$) are well observed.

\begin{figure}
\includegraphics[width=1\linewidth]{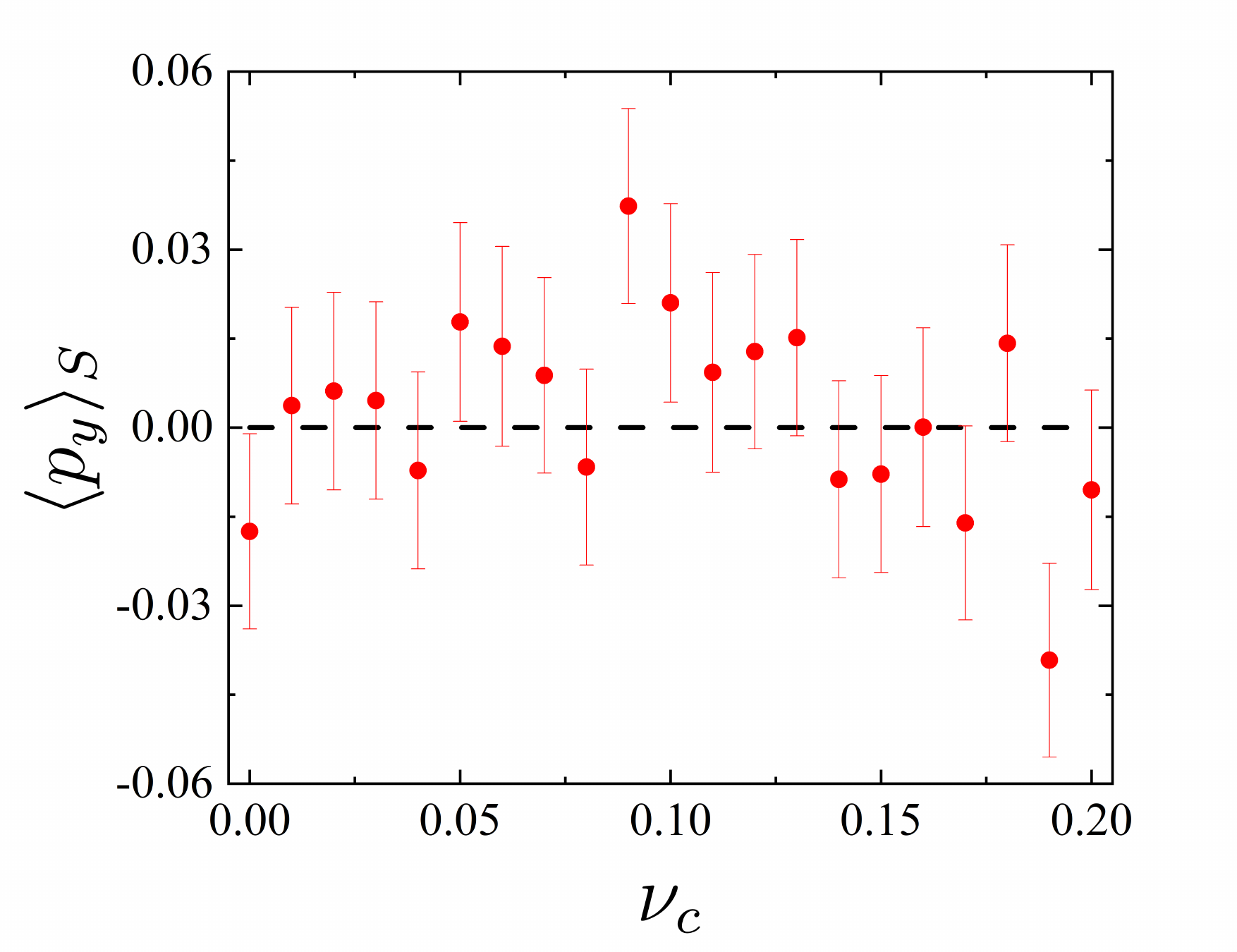}
\caption{Simulation results of $\langle p_y \rangle_S$ for spherical bacteria with different $\nu_c$. The solid circles and the error bars show the mean and the standard errors, respectively, over 4,000 independent simulation runs. Dashed line highlights $y=0$.}
\label{SphereS}
\end{figure}

Our results in Fig.\,\ref{Steady}a clearly demonstrate that $\langle p_y \rangle_S$ is not just a function of $\nu_c$, in which case the contour lines would have been vertical. Instead it is observed that the rheotactic behavior is weaker (stronger) at the lower left (upper right) corner where both $G$ and $\nu_c$ are small (big), indicating strong dependence on both $nu_c$ and $G$ (or equivalently bacterial aspect ratio $\alpha$). The dependence on $G$ is more significant in the lower right corner of Fig.\,\ref{Steady}a where the contour lines are almost horizontal, showing that bacterial aspect ratio is a crucial factor for bacterial rheotactic behavior. This conclusion is particularly true for the special case with $\alpha = 1 \left(G = 0\right)$ where the bacterium is effectively spherical: our results show $\langle p_y \rangle\approx0$ for all different choices of $\nu_c$ used in simulation, as illustrated in Fig.\,\ref{SphereS}. In other words, bacteria with $G=0$ can exhibit no rheotactic behavior even when $\nu_c \neq 0$.

\section{theoretical analysis and comparison}
To understand the results in previous section, here we would like to make a few approximations so that analytical results can emerge. In the first approximation, we assume that the coupling between the deterministic component and the stochastic component on the right hand side of Eq.~(\ref{eomOri}) are so small that the two components can be considered as independent. Thus bacterial orientational dynamics governed by Eq.~(\ref{eomOri}) can be obtained by considering the following two processes: the dynamics governed by the stochastic component only, and the dynamics governed by the deterministic component only.

The first process, i.e. the orientational dynamics under the influence of an angular diffusion, has been well studied with analytic prediction $d\langle \bm{p}(t) \rangle/dt = k_\mathrm{off}\langle \bm{p}(t) \rangle$ where $k_\mathrm{off}=-2D_r$.

\begin{figure}
\includegraphics[width=1\linewidth]{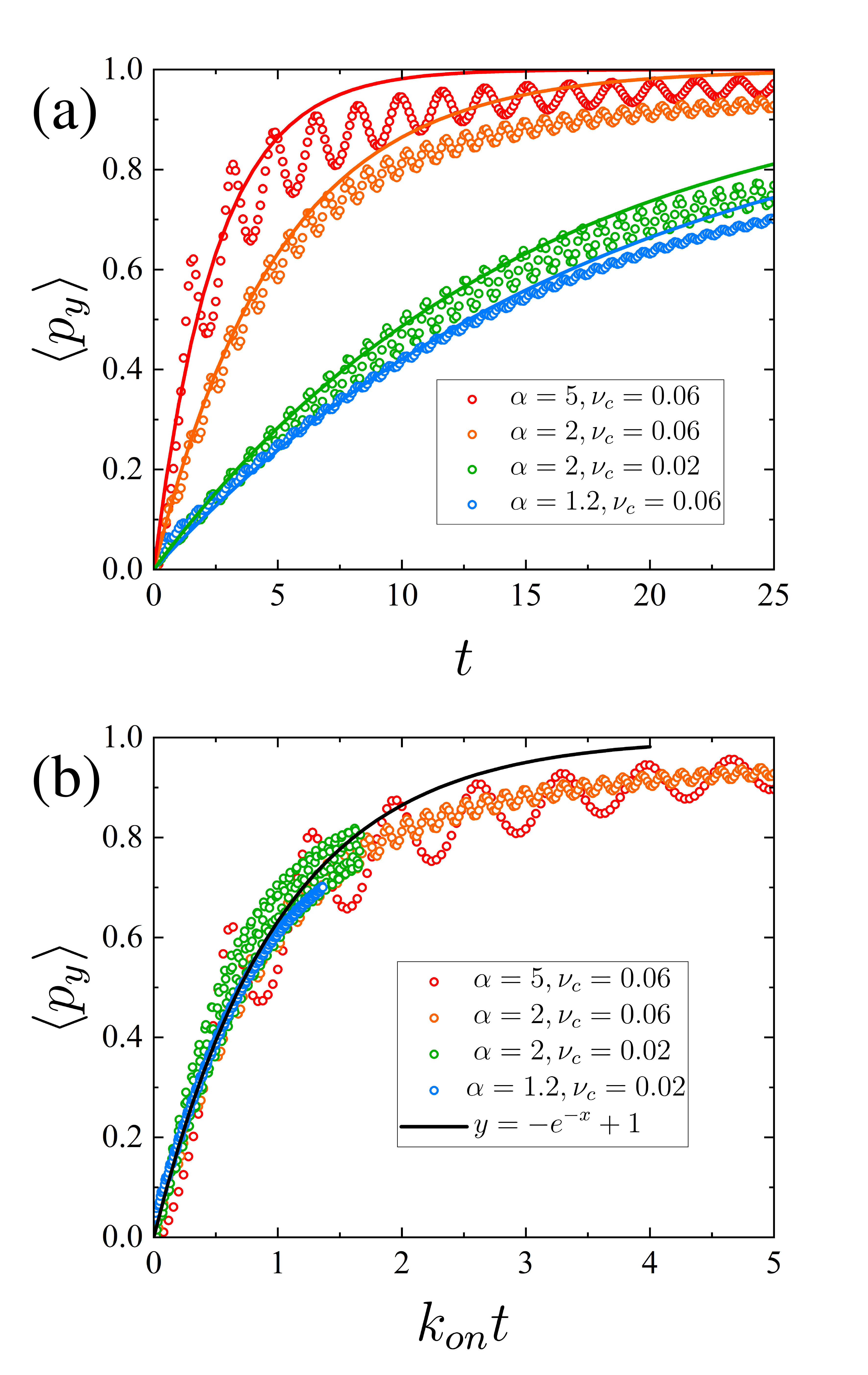}
\caption{For simple shearing flow with $\dot{\gamma}=10$, a comparison between theorectical predictions of Eq.~(\ref{OnProcess}) (solid lines) and numerically obtained $\langle p_y \rangle$ from the deterministeric dynamics described by Eq.~(\ref{azi}) and Eq.~(\ref{polar}) (open circles), as a function of $t$ (a) or $k_\mathrm{on}t$ where $k_\mathrm{on} \equiv \nu_c \dot{\gamma}(\alpha-1)/(\alpha+1)$ (b).}
\label{Temporal}
\end{figure}

\begin{figure}
\includegraphics[width=1\linewidth]{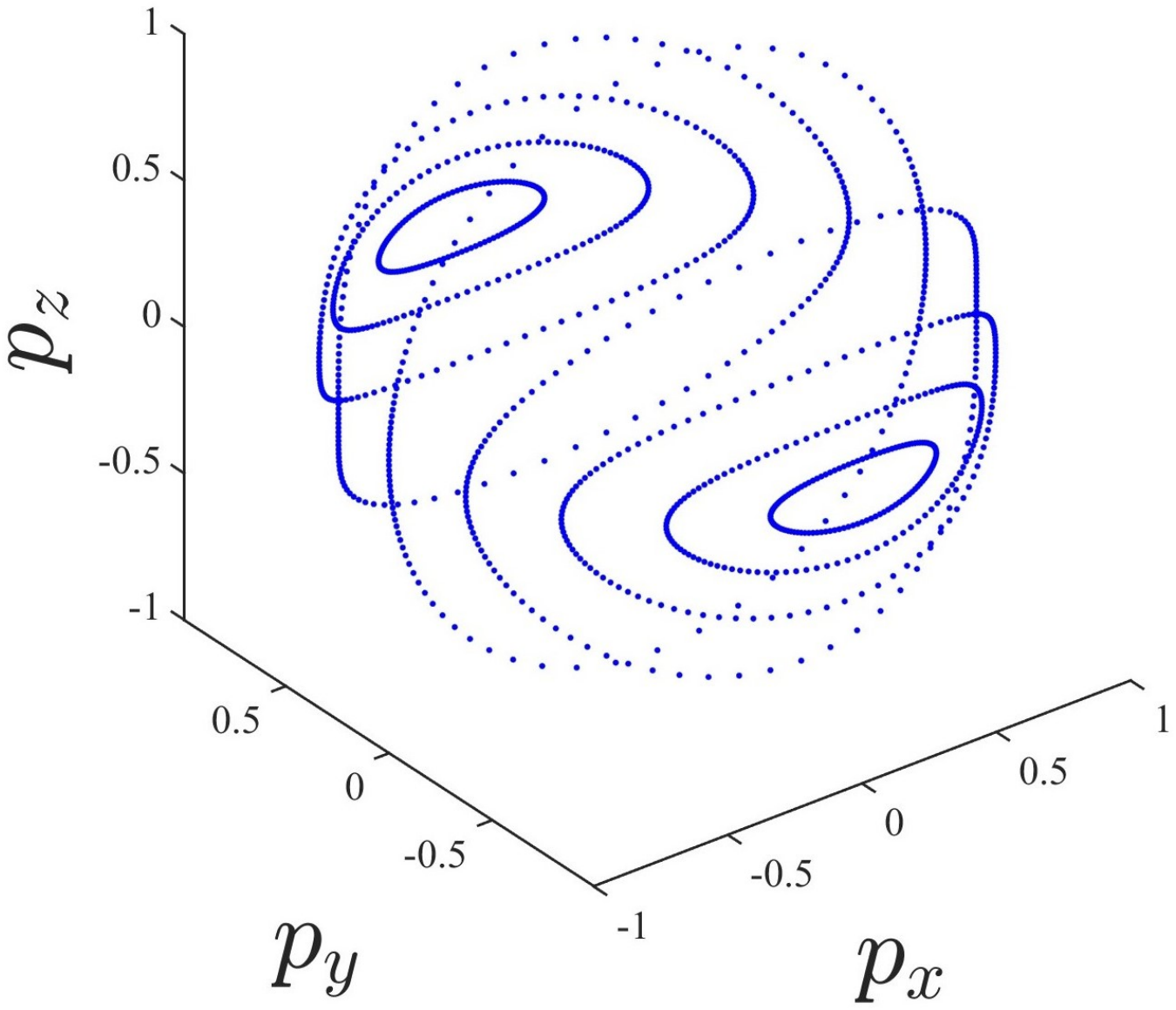}
\caption{A few sample Jeffery orbits. Each orbit is an enclosed cycle illustrated by dots separated by the same time step.}
\label{Jeffery}
\end{figure}

The second process, i.e. the orientational dynamics under the influence of the deterministic component only, can be described by Eq.~(\ref{azi}) and Eq.~(\ref{polar}). To investigate these deterministic dynamics, a few sets of simulations are performed in the same fashion as in Sec.\,III except here we ignore the noise terms and follow Eq.~(\ref{azi}) and Eq.~(\ref{polar}). From these simulations we obtained the orientational dynamics in terms of the temporal evolution of $\langle p_y \rangle$, with some examples illustrated as open symbols in Fig.\,\ref{Temporal}a. Two approximations are used to explain these data. The first one follows previous study \cite{Ronteix2022} to neglect the second term on the right hand side of Eq.~(\ref{azi}) so that changes in the azimuthal angle $\phi$ are purely caused by Jeffery reorientation. And the second one assumes that the azimuthal angle $\phi$ changes much faster than the polar angle $\theta$ so that $\theta$ barely changes over a period $T$ for $\phi$ that goes around a Jeffrey orbit as illustrated in Fig.\,\ref{Jeffery}, which is a good approximation for sphere-like swimmers and deteriorates at larger aspect ratios. With these two approximations, we can obtain the coarse-grained dynamics of $\theta$ by averaging the effect arisen from $\phi$ over the period $T$, in the form of:
\begin{equation}
\begin{aligned}
\dot{\theta}&=\nu_c\dot{\gamma} \cos{\theta}\dfrac{1}{T} \int_{0}^T \cos{2\phi}dt-G\dfrac{\dot{\gamma}}{4} \sin{2\theta} \dfrac{1}{T}\int_{0}^T\sin{2\phi}dt\\
&=\nu_c\dot{\gamma}\dfrac{\alpha-1}{\alpha+1} \cos{\theta}
\label{SingleTheta}
\end{aligned}
\end{equation}
Integrating Eq.~(\ref{SingleTheta}) leads to $\sin{\theta}=\frac{Ae^{Bt}-1}{Ae^{Bt}+1}$, where $A=(1+\sin{\theta_0})/(1-\sin{\theta_0})$ is a constant determined by the initial condition $\theta(t=0)=\theta_0$, and $B=2\nu_c\dot{\gamma}(\alpha-1)/(\alpha+1)$. Assuming evenly distributed $\sin{\theta}$ at the initial condition, we get $\langle \sin{\theta} \rangle = \int_{-1}^{1} \frac{Ae^{Bt}-1}{Ae^{Bt}+1}d{(\sin{\theta_0})}$, which approximately leads to 
\begin{equation}
    \langle p_y(t) \rangle\equiv\langle \sin{\theta} \rangle =-e^{-k_\mathrm{on}t}+1
    \label{OnProcess}
\end{equation}
where $k_\mathrm{on} \equiv \nu_c \dot{\gamma}(\alpha-1)/(\alpha+1)$. Predictions of Eq.~(\ref{OnProcess}) are illustrated as solid lines in Fig.\,\ref{Temporal}a, which are in good agreement with simulation results (open symbols) for sphere-like bacteria with $\alpha\approx 1$. For bacteria with larger aspect ratio, our simulations data show increasingly larger oscillations in the temporal evolution of $\langle p_y \rangle$ beyond the exponential decay predicted by Eq.~(\ref{OnProcess}). The difference is expected, as our approximation for time scale separation between the dynamics of $\phi$ and $\theta$ is only accurate for sphere-like bacterium with $\alpha\approx 1$, and deteriorates at larger $\alpha$. That is, for bacteria with large aspect ratio changes in $\phi$ can no longer be considered as fast, and the coupling between the $\phi$ and $\theta$ directly leads to the oscillatory behavior in the temporal evolution of $\langle p_y \rangle$ observed in simulation. But overall the exponential decay does semi-quantitatively capture the trend, as our numerically observed $\langle p_y \rangle$ for bacteria with different $\alpha$ collapse to a master curve $y=1-e^{-x}$ when plotted against $k_\mathrm{on}t$, as illustrated in Fig.\,\ref{Temporal}b.

Combining the first process where $d\langle p_y \rangle/dt=k_\mathrm{off}\langle p_y \rangle$ and the second process where $d\langle p_y \rangle/dt=k_\mathrm{on}(1-\langle p_y \rangle)$, we can obtain:
\begin{equation}
\dfrac{d\langle p_y (t) \rangle}{dt}=k_\mathrm{on}(1-\langle p_y (t) \rangle)+k_\mathrm{off}\langle p_y (t) \rangle.
\label{ratePy}
\end{equation}
With initial condition $\langle p_y (t=0) \rangle=0$, bacterial orientational dynamics can be obtained as
\begin{equation}
\langle p_y (t) \rangle=\dfrac{k_\mathrm{on}}{k_\mathrm{on}-k_\mathrm{off}}\left(1-e^{-(k_\mathrm{on}-k_\mathrm{off})t}\right).
\label{dynaPy}
\end{equation}
Taking the $t \to \infty$ limit, Eq.~(\ref{dynaPy}) describes the steady state as
\begin{equation}
\langle p_y \rangle_S = k_\mathrm{on}/(k_\mathrm{on}-k_\mathrm{off}),
\label{SteadyState}
\end{equation}
where we have $k_\mathrm{on}\equiv\nu_c\dot{\gamma}(\alpha-1)/(\alpha+1)$, and $k_\mathrm{off}\equiv -2D_r$.

From Eqs.\,\ref{dynaPy} and \ref{SteadyState}, it is clear that bacterial orientational dynamics is controlled by two dimensionless parameters: the rotational Peclet number $Pe\equiv\dot{\gamma}/D_r$ characterizing the ratio of external drive to thermal noise, and an intrinsic factor of bacterial dimensions $\nu_c(\alpha-1)/(\alpha+1)$. In comparison to existing theories \cite{jing2020} which describe bacterial chirality-induced rheotaxis in parameter $\nu_c\dot{\gamma}/D_r$, our study here point out the contribution of bacterial aspect ratio $\alpha$, which manifests in the form of an independent multiplier $(\alpha-1)/(\alpha+1)$. This multiplier is in particular important for sphere-like microswimmers with $\alpha\approx1$, and becomes less significant as $\alpha$ increases, eventually reduces to $1$ in the $\alpha\to\infty$ limit for needle-like microswimmers. For the special case with $\alpha=1$ where the bacterium is effectively spherical, Eq.~(\ref{SteadyState}) predicts $\langle p_y \rangle_S = 0$ irrespective of $\nu_c$, which is in good agreement with our simulation results (Fig.\,\ref{SphereS}).

\begin{figure}
\includegraphics[width=1\linewidth]{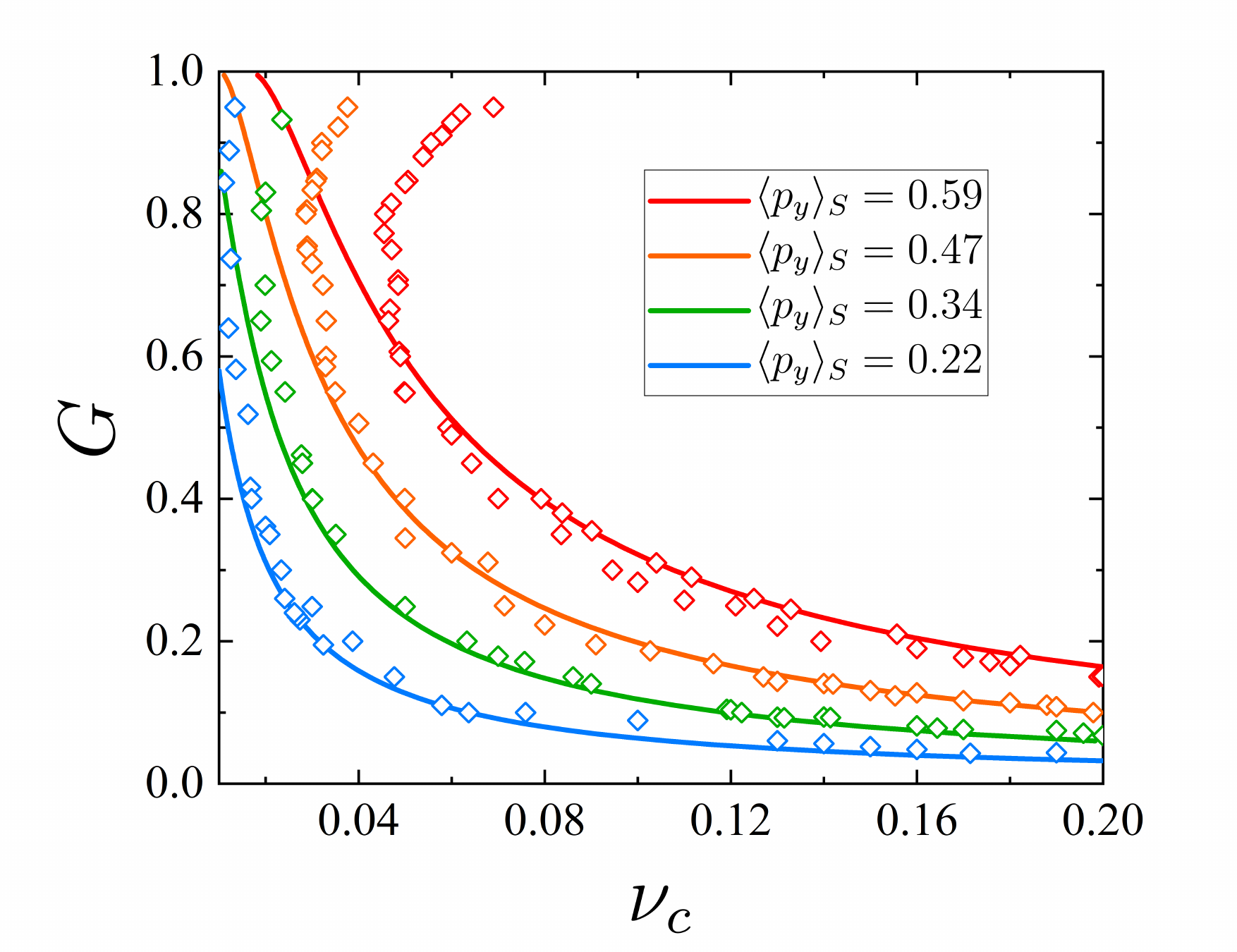}
\caption{A few sample contour lines extracted from Fig.\,\ref{Steady}, with $\langle p_y \rangle_S$ being different constants. Results from numerical simulations (open spades) are shown in comparison with predictions by Eq.~(\ref{SteadyState}) with $k_\mathrm{on}\equiv\nu_c\dot{\gamma}(\alpha-1)/(\alpha+1)$ (solid lines).}
\label{Contour}
\end{figure}

For a more general comparison, in Fig.\,\ref{Steady}b we show theoretical predictions by Eq.~(\ref{SteadyState}) in a contour map that is in the same fashion as our simulation results in Fig.\,\ref{Steady}a. Again, for sphere-like bacteria with $G \ll 1$ we see very good agreement, and the difference increases at larger $G$ and becomes noticeable at around $G=0.6$ (or equivalently $\alpha=2$). To make a more quantitative comparison, in Fig.\,\ref{Contour} a few sample contour lines of $\langle p_y \rangle_S$ predicted by Eq.~(\ref{SteadyState}) are illustrated as solid lines, while the corresponding contour lines obtained from our numerical simulations in section III based on Eq.~(\ref{eomOri}) are illustrated as open squares. For $0 \leq G \leq 0.6$ (or equivalently $1 \leq \alpha \leq 2$), our numerical results are in quantitative agreement with theoretical predictions, supporting our conclusion that for sphere-like bacteria the rheotactic behavior is strongly dependent on bacterial aspect ratio $\alpha$ (or equivalently $G$) otherwise the contour line would have been vertical. For $G > 0.6$ (or equivalently $\alpha > 2$), there is noticeable differences between numerical simulations and theoretical predictions in Fig.\,\ref{Contour}. We believe that the differences arise from the fact changes in $\phi$ can no longer be considered as fast and our approximation of the time scale separation is no longer valid.

\begin{figure}
\includegraphics[width=1\linewidth]{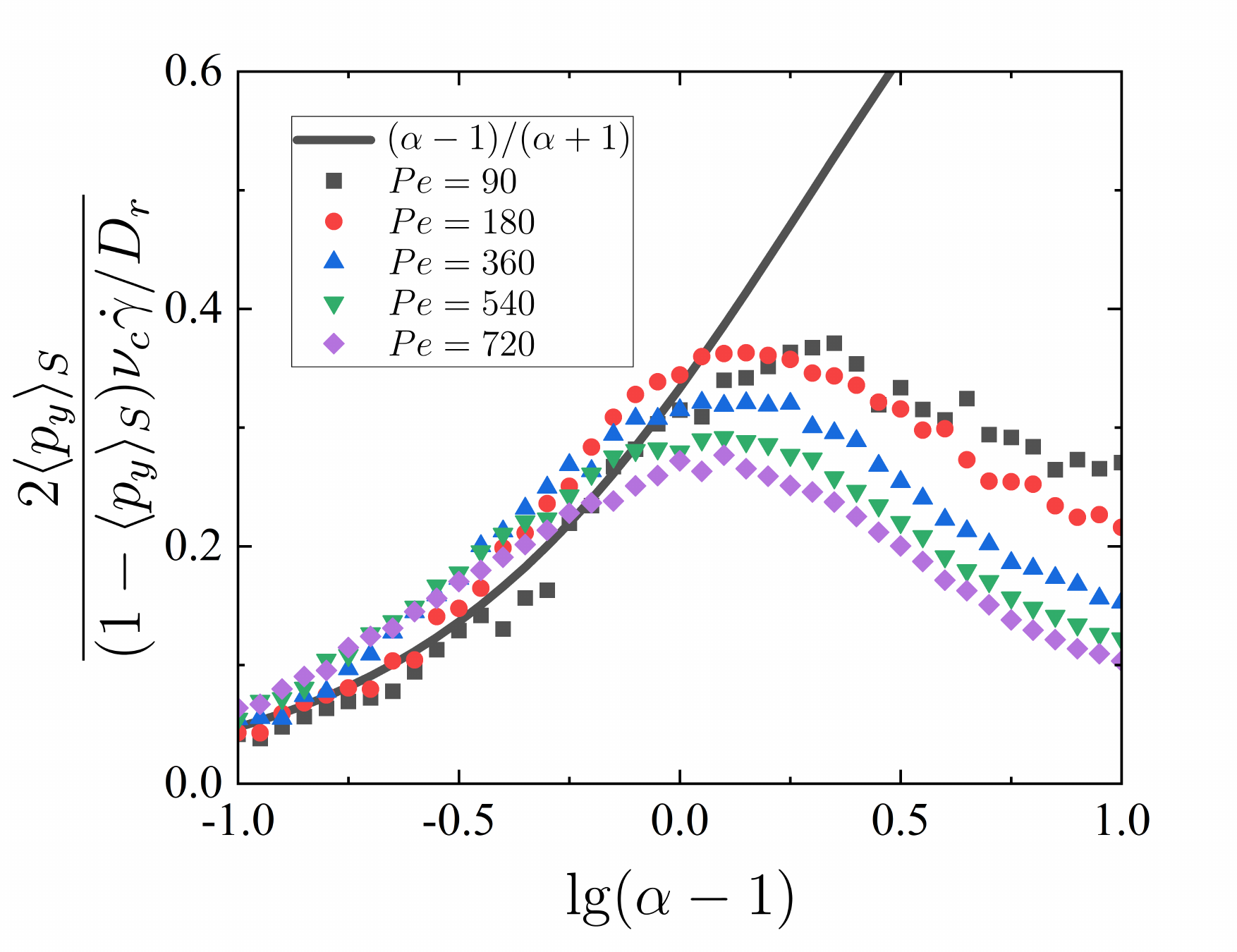}
\caption{The contribution from cell aspect ratio $\alpha$ can be included through an independent multiplier, which is predicted to be a simple function of $\alpha$ (black line) and can be extracted from our simulation data (symbols).}
\label{AlphaEffect}
\end{figure}

To further investigate how general the theoretical prediction Eq.~(\ref{SteadyState}) applies, we performed more numerical simulations for simple shearing flows with other choices of $\dot{\gamma}$. For a better comparison between our simulation results and the theoretical prediction, we rewrite Eq.~(\ref{SteadyState}) to highlight the bacterial aspect ratio contribution:
\begin{equation}
\frac{\alpha-1}{\alpha+1}=\frac{2\langle p_y \rangle_S}{(1-\langle p_y \rangle_S)\nu_c\dot{\gamma}/D_r}.
\label{multiplier}
\end{equation}
The right hand side of Eq.~(\ref{multiplier}) can be obtained from numerical simulations for different bacterial aspect ratio $\alpha$, at different simple shearing flows $\dot{\gamma}$, as illustrated in Fig.\,\ref{AlphaEffect} as symbols. The left hand side of Eq.~(\ref{multiplier}) is plotted in the same figure as the solid black line. The comparison further validated our conclusion: for sphere-like bacterium ($1 \leq \alpha \leq 2$) theoretical prediction is in quantitative agreement with numerical simulations that bacterial rheotactic behavior depends strongly on bacterial aspect ratio through a multiplier $(\alpha-1)/(\alpha+1)$; for needle-like bacteria ($\alpha > 2$) theoretical prediction Eq.~(\ref{SteadyState}) or Eq.~(\ref{multiplier}) is no longer accurate as the slowing down of dynamics in $\phi$ makes the time scale separation approximation no longer valid.

\section{Discussion and conclusion}
In this work we have investigated the bacterial rheotactic response to external flows. Our numerical results clearly show that bacterial aspect ratio as an independent parameter is a crucial factor, especially for sphere-like bacteria with aspect ratio $\alpha\approx 1$, where for the limiting case of an effective sphere with $\alpha=1$ no rheotactic behavior exists irrespective bacterial chirality $\nu_c$. An analytic explanation is provided (Eq.~(\ref{SteadyState})), where we modify the key parameter $\nu_c\dot{\gamma}/D_r$ by introducing an independent multiplier $(\alpha-1)/(\alpha+1)$, based on two major approximations. The first approximation assumes that the external drive and the stochastic decay are two independent processes in the orientational dynamics, which is generally believed to be true. The second approximation assumes that changes in azimuthal angle $\phi$ is much faster than changes in polar angle $\theta$, allowing us to average out the quick dynamics in $\phi$ for the slow dynamics in $\theta$. Our results show that this approximation is reasonable for sphere-like bacteria with $1 \leq \alpha \leq 2$, but deteriorates as $\alpha$ increases beyond 2.

\begin{acknowledgments}
This work is supported by NSFC $\# 12474191$, $\# 11974038$ and $\# \mathrm{U}2230402$. We also acknowledge the computational support from the Beijing Computational Science Research Center.
\end{acknowledgments}

\appendix

\section{Relationship Between Chiral Strength Factor and Cell Aspect Ratio}
\label{appendix A}
A bacterium is modelled as a body consisting of a prolate spheroid head attached to a left-handed rigid helix. The helix in shear flow experiences a drift velocity $\boldsymbol{v}$ \cite{marcos2009}, which is the source of the rheotactic torque \cite{marcos2012}. We use $\boldsymbol{p}$ to represent the orientation of the bacterium. We introduce the laboratory frame $\{\hat{\boldsymbol{x}}, \hat{\boldsymbol{y}}, \hat{\boldsymbol{z}}\}$ and impose an external shear flow $\boldsymbol{v}_f=\dot{\gamma}z\hat{\boldsymbol{x}}$. The angular velocity of the helix is gooed approximation of the Jeffery reorientation rate \cite{mathijssen2019}. Thus, the bacterium has an effective aspect ratio $\alpha = a/b$ for the Jeffery reorientation rate. Here, the long axis $a$ and short axis $b$ refer to the effective long and short axes, respectively. Disregarding the helix's drift velocity, the hydrodynamic torque exerted on the bacterium is expressed as \cite{kim2013}
\begin{multline}
\boldsymbol{T}_J=8\pi\mu a^3\left(X^C\boldsymbol{p}\boldsymbol{p}+Y^C\left(\boldsymbol{\delta}-\boldsymbol{p}\boldsymbol{p}\right)\right)\cdot\left(\boldsymbol{\Omega}^\infty-\boldsymbol{\Omega}\right)\\
-8\pi\mu a^3Y^H\left(\boldsymbol{E}^\infty\cdot\boldsymbol{p}\right)\times\boldsymbol{p}
\label{1b}
\end{multline}
with $\mu$ is viscosity of flow, where $\boldsymbol{\Omega}^\infty$ and $\boldsymbol{E}^\infty$ represent the rate-of-rotation and rate-of-strain of the external shear flow. $X^C$, $Y^C$ and $Y^H$ are scalar resistance functions, only depend on the eccentricity of generating ellipse $\chi=\left(a^2-b^2\right)^{1/2}/a$. $\boldsymbol{\Omega}$ is the rate of rotation of the bacteria. If the bacterium is torque-free in a shear flow, we can derive the well-known Jeffery reorientation rate $\boldsymbol{\Omega}^J$ as follows \cite{kim2013,zottl2013}
\begin{equation}
    \boldsymbol{\Omega}^\infty-\boldsymbol{\Omega}^J=\dfrac{Y^H}{Y^C}\left(\boldsymbol{E}^\infty\cdot\boldsymbol{p}\right)\times\boldsymbol{p}
\end{equation}
where $Y^H/Y^C=G$. The head of swimming bacteria acts as an anchor for drifting flagellar bundle \cite{marcos2012}. The drift velocity $\boldsymbol{v}$ of the helix in shear flow is written as \cite{mathijssen2019} 
\begin{equation}
\begin{aligned}
v_x&=2k_{\beta}\dot{\gamma}p_x p_y\\
v_y&=2k_{\beta}\dot{\gamma}\left(p_z^2 -p_x^2\right)\\
v_z&=-2k_{\beta}\dot{\gamma}p_y p_z
\end{aligned}
\end{equation}
where the prefactor $k_{\beta}$ only depends on the helix geometry. The torque on the bacterium by the drifting of helix can be written as
\begin{equation}
    \boldsymbol{T}_C=-\mu k_\alpha \boldsymbol{p}\times \boldsymbol{v}
\end{equation}
Where $k_\alpha$ is related to the helix geometry and size of the prolate spheroid head. This hydrodynamic torque induced by the drift velocity of the helix will balance the hydrodynamic torque of equation \eqref{1b}. As a result, a chirality-induced orientation rate, denoted as $\boldsymbol{\Omega}^C$, emerges. Rewrite the rate of rotation of the bacterium $\boldsymbol{\Omega}$ as $\boldsymbol{\Omega}^J+\boldsymbol{\Omega}^C$, we can write the torque-free relationship for the entire bacterium as
\begin{equation}    
-\mu k_\alpha \boldsymbol{p}\times \boldsymbol{v}-8\pi\mu a^3\left(X^C\boldsymbol{p}\boldsymbol{p}+Y^C\left(\boldsymbol{\delta}-\boldsymbol{p}\boldsymbol{p}\right)\right)\cdot\boldsymbol{\Omega}^C=0
\end{equation}
The expression for the chirality-induced reorientation rate $\boldsymbol{\Omega}^C$ is derived as
\begin{equation}
    \boldsymbol{\Omega}^C=-\dfrac{1}{8\pi}\dfrac{k_\alpha}{a^3}\left(Y^C\right)^{-1}\left(\boldsymbol{p}\times\boldsymbol{v}\right)
\end{equation}
where 
\begin{equation}
\begin{aligned}
Y^C=\dfrac{4}{3}\chi^3\left(2-\chi^2\right)
\left(-2\chi+\left(1+\chi^2\right)ln\left(\dfrac{1+\chi}{1-\chi}\right)\right)^{-1}
\end{aligned}
\end{equation}
The chiral strength factor $\nu_c=k_\alpha k_\beta(4\pi a^3)^{-1}\left(Y^C\right)^{-1}$. Typically, the bacterial head is much larger in volume than the tail, so the helix geometry has a minimal effect on the effective aspect ratio $\alpha$. Here, we approximate that the helix geometry is independent of $\alpha$. Since $k_\alpha$ and $k_\beta$ are independent of $\alpha$, when size of the bacterium $a$ remains constant, the relationship between $\nu_c$ and $\alpha$ is effectively equivalent to the relationship between $(Y^C)^{-1}$ and $\alpha$.

\bibliography{ShapeMatters}

\end{document}